\documentclass[nofootinbib,twocolumn,prl,aps]{revtex4}
\usepackage{graphicx}

\begin{document}
\title{Realistic clocks, universal decoherence and 
the black hole information paradox}
\author{Rodolfo Gambini$^{1}$, Rafael A. Porto$^{2}$ and 
Jorge Pullin$^{3}$}
\affiliation {1. Instituto de F\'{\i}sica, Facultad de Ciencias,
Igu\'a 4225, esq. Mataojo, Montevideo, Uruguay. \\ 
2. Department of Physics, Carnegie Mellon University, Pittsburgh,
PA 15213\\
3. Department of Physics and Astronomy, Louisiana State University,
Baton Rouge, LA 70803-4001}
\date{June 26th 2004}

\begin{abstract}
Ordinary quantum mechanics is formulated on the basis of the existence
of an ideal  classical clock external to the system under
study. This is clearly an idealization. As emphasized originally by
Salecker and Wigner and more recently by other authors,
there exist limits in nature to how ``classical'' even the
best possible clock can be. When one introduces realistic clocks,
quantum mechanics ceases to be unitary  and a fundamental
mechanism of decoherence of quantum states arises. We estimate the
rate of universal loss of unitarity using optimal realistic clocks.
In particular we observe that the rate is rapid enough to eliminate
the black hole information puzzle: all information is lost through the
fundamental decoherence before the black hole can evaporate.
This improves on a previous calculation we presented with a
sub-optimal clock in which only part of the information was lost
by the time of evaporation.
\end{abstract}
\maketitle

The question of how classical can an optimal clock be was first
considered by Salecker and Wigner \cite{Wigner}, who noted that in
order to measure time with increasing accuracy one needs increasing
quantities of energy.  Specifically, they construct an elementary
clock consisting of a photon bouncing between two mirrors separated by
a distance $\ell$.  The clock ``ticks'' every time the photon bounces
on one of the mirrors \cite{Wigner}.  By the time the photon returns
after one bounce, the wavepacket of the mirror has spread, leading to
a bound in the accuracy of the time measurement of $\left(\delta T
\right)^2 \gtrsim \hbar T_{\rm max} /(m c^2)$ with $m$ the mass of the
clock and $c$ the speed of light and $T_{\rm max}$ is the maximum
interval of time one attempts to measure. The measurement becomes more
inaccurate the longer the time to be measured, and the smaller the
mass of the clock. Amelino-Camelia and Ng and Van Dam further
elaborated on this idea \cite{AcNg} by noticing that a fundamental
limit exists on how accurate a clock can be: if one needs more
accuracy, the energetic demands are so high that the clock collapses
into a black hole (the size of the clock cannot be increased to 
prevent the collapse, since it would imply losing accuracy). 
In fact, a black hole is the most accurate clock
available for a given mass. A simple way of viewing the black hole as
a clock is to recall that when excited, black holes behave like a
(damped) oscillator. The fundamental frequency is inversely
proportional to the mass of the hole, and therefore the resolution of
the black hole as a clock is proportional to its mass.  Moreover,
since Hawking \cite{Ha} showed that black holes evaporate due to
particle production, one has a maximum possible time that can be
measured by a black hole clock. If we take this time to be the black
hole evaporation time, the inequality listed above is satisfied as an
equality. Therefore if one wishes to measure time intervals smaller
than a certain value $T_{\rm max}$, the optimal clock is a black hole
with lifetime (at least) $T_{\rm max}$, a bigger black hole will be
less accurate, a smaller one will evaporate too fast to operate as a
clock. The fundamental accuracy with which one can measure a time
$T_{\rm max}$ is therefore determined by the lifetime of the black
hole and is given by
\begin{equation}
\delta T \sim t_P \sqrt[3]{T_{\rm max}/t_P} \label{evap}
\end{equation}
where $t_P$ is Planck's time and from now on we choose units where
$\hbar=c=1$.

In order to do quantum mechanics with realistic clocks, one has to
include the clock as part of the system under study. A suitable
construction has been proposed by Page and Wootters \cite{PaWo} and a
recent reanalysis is present in the paper by Dolby \cite{Dolby}. What
one does it to compute probabilities for quantities of the system
under study conditional on the quantities describing the clock taking
given values. If the clock behaves semiclassically, the resulting
probabilities satisfy approximately a Schr\"odinger equation. However,
since the clock can never behave entirely classically, there will be
corrections, at least if one wishes to recover Schr\"odinger's
equation at a leading order \cite{Marolf}. We have estimated the type
of corrections in reference \cite{GaPoPu} in the context of a discrete
theory \cite{GaPoPu0} but the construction can also be applied to the continuum
case. In particular, the corrections imply that the quantum states do
not evolve unitarily.  Notice that the argument is based on ordinary
(unitary) quantum mechanics, we are just recasting the theory in terms
of a realistic clocks and this is the root of the loss of unitarity.
The magnitude of the loss of unitarity is characterized by a function
with units of time that is associated with how accurate the clock one
considers is with respect to an ideal classical clock.

We briefly recount the derivation of the decoherence formula from
reference \cite{GaPoPu}. We consider a system described by a variable
$X$ and a clock described by a variable $T$. Both variables are
treated quantum mechanically and evolve according to Schr\"odinger's
theory with respect to an ideal time $t$. We can start the system in
an optimal quantum state for the clock, in which the probability
density for the variable $T$ has the shape of a Dirac delta centered
at $T=t_0$. However, upon evolution, the probability distribution will
spread and there will be several likely values of $T$ for a given
instant of the ideal clock $t$.  We would do quantum mechanics by
computing the conditional probability asking the question what is the
probability of the variable $X$ taking a given value $X_0$, when the
clock variable $T$ takes the value $T_0$ (if $T$ and $X$ have 
continuous spectra, we should recast the question in terms of 
small intervals). Since for later times there
will be several values of the external time $t$ that correspond to the
value of the clock $T_0$, the resulting probability will be a
superposition of ordinary Schr\"odinger probabilities. The latter are
evolved unitarily, the former is therefore not. Detailed calculations
in reference \cite{GaPoPu} show that one can approximate the evolution
(provided the clock is reasonably classical) by a Lindblad type
evolution,
\begin{equation}
{\partial \rho \over \partial T} = -i[H,\rho]- \sigma(T) [H,[H,\rho]]
\label{lindblad}
\end{equation}
where $\rho$ is the density matrix describing the system under
study (without the clock) and $\sigma(T)$ is a measure of the
rate of spread of the probability distribution of the clock time 
as a function of the ideal time. Specifically, if we assume
the probability distribution is a Gaussian of spread $\delta T$,
$\sigma(T)=\partial \left(\delta T\right)^2/\partial T$.

Since we have argued what an optimal clock is, we can now estimate
what is the minimum rate of non-unitarity that one can expect from
quantum mechanics in the real world by providing a concrete model for
the spread $\sigma(T)$. Notice that this effect is fundamental, it
affects all physical systems and cannot be eliminated. In particular,
it does not depend on any interaction of the clock with the system.
Quantum systems can decohere due to other effects, and in many
practical applications these will operate much faster than the
fundamental effect we discuss here \cite{GaPoPu}. The latter is
nevertheless ever present. The formula we get starting from
(\ref{evap}) for $\sigma(T)$ for an optimal clock is given by,
$\sigma(T)= \left({t_{P}}\over{T_{\rm max}-T}\right)^{1/3}t_{P}$,
where $T_{\rm max}$ is the length of time we wish to measure and we
take it to coincide with the evaporation time of the black hole.

We now turn our attention to the black hole information paradox.
Simply stated (for a review see \cite{GiTh}) the paradox goes as
follows: take a pure quantum state and collapse it into a black
hole. Let the black hole evaporate. The end state is the outgoing
thermal radiation, that is, a mixed state. In ordinary quantum
mechanics, since evolution has to be unitary, a pure state cannot
evolve into a mixed state, hence the puzzle. As we argued above, if
one uses realistic clocks in ordinary quantum mechanics, pure states
can evolve into mixed states. There is therefore the possibility that
the collapse into a black hole and subsequent evaporation of a pure
quantum state may not constitute a puzzle. The requirement is that the
fundamental decoherence, that would turn the pure state into a mixed
one anyway, operate fast enough to occur before the black hole
evaporates entirely. We will now show that this is the case.  In a
previous paper we analyzed this problem using a sub-optimal clock
\cite{GaPoPusuboptimal}. The current calculation yield a better
picture in the sense that it implies that {\em all} information is
lost by the time the black hole evaporates, which was not the case
with the sub-optimal clock.

We need to make a quantum model of the black hole in order to study
its decoherence. Here we will make a very primitive model. We assume
the black hole horizon's area (or equivalently its energy) is
quantized. This is usually assumed in quantum black hole studies and
in particular it is predicted by loop quantum gravity. We choose a
basis of states for the black hole labeled by the energy (area).
The problem has some resemblance to the problem of an atom that
is in an excited state and emits radiation to reach its fundamental
state. If one considers the physical system under study to be the
atom plus the radiation field, its evolution is unitary. One would
expect a similar situation to hold for the black hole interacting
with the gravitational and matter fields surrounding it. Here is
where the paradox lies, since the evaporation process leads to loss
of unitarity for the total system. Our model will include information
about the black hole and the surrounding fields such that it
starts its evolution in a pure state, and we will study its 
evolution according to equation (\ref{lindblad}). We consider
the system as described by a density matrix,
\begin{equation}
\rho = \sum_{ab} \rho_{ab} |E(T)+\epsilon_a,E_0-E(T)> 
<E(T)+\epsilon_b,E_0-E(T)|, 
\end{equation}
where the first entry in the bra (ket) represents the energy
of the black hole at instant $T$, which changes with time in an
adiabatic fashion, the constant $E_0$ represents
the mean value of the 
total energy of the system (which is conserved) and 
$E_0-E(T)$ is the energy of the field at instant $T$. We consider
the state to be a superposition of states of the black hole that
differ in energy from $E(T)$ by $\epsilon_{a}$. To simplify the
analysis we consider only a pair of levels of energy that are
separated by an energy proportional to the temperature,
as one would expect for an evaporating hole.
Concretely, the
characteristic frequency for this energy is given by
\begin{equation}
\omega_{12}(T) = {1 \over \left(8 \pi\right)^2 
t_P }\left(t_P\over T_{\rm max}-T\right)^{1/3}
\end{equation}
with $T_{\rm max}$ the lifetime of the black hole (how long it takes to
evaporate) and the subscript $12$ denotes that it is the transition
frequency between the two states of the system.  Although this model
sounds simple-minded it just underlies the robustness of the
calculation: it just needs that the black hole have discrete energy
levels characterized by a separation determined by the temperature of
the black hole.  It is general enough to be implemented either 
assuming the Bekenstein spectrum of area or the spectrum stemming
from loop quantum gravity \cite{BaCaRo}.
We assume that we start with the black hole in a pure
state which is a superposition of different energy eigenstates
(there is no reason to assume that the black hole is  exactly in 
an energy eigenstate, which would imply a stationary state with no
radiation being emitted; as soon as one takes into account the
broadening of lines due to interaction one has to consider a
superposition of states within the same broadened level with a
time dependent separation with a similar behavior).
Therefore the density matrix has off-diagonal
elements.  One now needs to write equation (\ref{lindblad})
in the simplified energy basis we chose and one can immediately
integrate it to yield,
\begin{equation}
\log\left(\rho_{12}(T)\over \rho_{12}(0)\right) 
= i \int_0^T \omega_{12}(T') dT'
+{1 \over (8\pi)^2} \log\left(T_{\rm max}-T\over T_{\rm max}\right).
\end{equation}
We therefore see that when time reaches the evaporation time 
$T=T_{\max}$,
the density matrix element vanishes, i.e. the state has decohered
completely. Therefore there is no information puzzle to be contended
with.

The result presented above is remarkable for being able to erase
completely the information before evaporation.  On the other hand, it
is clear that we have taken a very crude model for the black hole and
a more detailed calculation is needed before one can completely write
off the black hole information puzzle, but the present calculation
provides good hope that the problem can indeed be solved. A realistic
calculation seems somewhat beyond the state of the art. For instance,
it is clear that the calculation should model quantum mechanically the
black hole but also the fields it interacts with in a detailed way
in the context of a theory of quantum gravity.

Returning to other physical systems, a similar calculation with a 
two level system yields that the level of fundamental decoherence
is,
\begin{equation}
\log\left(\rho_{12}(T)\over \rho_{12}(0)\right)= -{3 \over 2} 
t_P^{(4/3)} T^{(2/3)}
\omega_{12}^2.
\end{equation}
The effect is too small to be observed in the lab, unless one can
construct a system with a significant energy difference between
the two levels. The most promising candidate systems would be given
by systems of ``Schr\"odinger cat'' type. Bose--Einstein condensates
could in some future provide a system where the effect could be 
close to observability \cite{GaPoPu}.

Summarizing, we have shown that unitarity in quantum mechanics 
only holds when describing the theory in terms of a perfect idealized
clocks. If one uses realistic clocks loss of unitarity is introduced.
We have estimated a minimum level of loss of unitarity based on 
constructing the most accurate clocks possible. The loss of unitarity
is universal, affecting all physical phenomena. We have shown that
although the effect is very small, it may be important enough to
avoid the black hole information puzzle.

This work was
supported by grant NSF-PHY0244335 and funds from the Horace Hearne
Jr. Institute for Theoretical Physics and the International Centre
for Theoretical Physics (Trieste, Italy).

\end{document}